\documentclass[12pt]{article}
\usepackage{amsmath}
\usepackage{amssymb}
\usepackage{amsfonts}
\usepackage{bbold}
\usepackage{graphicx,psfrag,epsf}
\usepackage{enumerate}
\usepackage{natbib}
\usepackage{url} 
\usepackage[utf8]{inputenc}
\usepackage[OT4]{fontenc}

\newcommand{\bz}{\boldsymbol{z}}
\newcommand{\bZ}{\boldsymbol{Z}}
 \newcommand{\bbeta}{\boldsymbol{\beta}}
 \newcommand{\beeta}{\boldsymbol{\eta}}
 \newcommand{\bdelta}{\boldsymbol{\delta}}
\newcommand{\blind}{1}

\begin{document}

\def\spacingset#1{\renewcommand{\baselinestretch}%
{#1}\small\normalsize} \spacingset{1}


\if1\blind
{
  \title{\bf Estimation of the Size of Informal Employment Based on Administrative Records with Non-ignorable Selection Mechanism}
  \author{Beręsewicz Maciej\hspace{.2cm}\\
    \small{Department of Statistics, Poznań University of Economics and Business, Poland}\\
    \small{Centre for Small Area Estimation, Statistical Office in Poznań, Poland}\\
    \small{ORCID  0000-0002-8281-4301}\\
    Nikulin Dagmara \\
    \small{Faculty of Management and Economics, Gdańsk University of Technology, Poland}\\
    \small{ORCID 0000-0002-0534-4553}}
   \maketitle
} 
\fi
\if0\blind
{
  \bigskip
  \bigskip
  \bigskip
  \begin{center}
    {\LARGE\bf Estimation of the Size of Informal Employment Based on Administrative Records with a~Non-ignorable Selection Mechanism}
\end{center}
  \medskip
} \fi

\begin{abstract}
In this study we used company level administrative data from the National Labour Inspectorate and The Polish Social Insurance Institution in order to estimate the prevalence of informal employment in Poland in 2016. Since the selection mechanism is~non-ignorable we employed a~generalization of Heckman's sample selection model assuming non-Gaussian correlation of errors and clustering by incorporation of random effects. We found that 5.7\% (4.6\%, 7.1\%; 95\% CI ) of registered enterprises in Poland, to some extent, take advantage of the informal labour force.  Our study exemplifies a~new approach to measuring informal  employment,  which  can  be  implemented  in  other  countries. It also contributes to the existing literature by providing, to the best of our knowledge, the first estimates of informal employment at the level of companies based solely on administrative data.
\end{abstract}

\noindent%
\vfill

\newpage
\spacingset{1.45} 
\section{Introduction} 

Empirical research on the shadow economy is~extremely difficult, given limited knowledge about the characteristics of informal workers. Those who work informally want to protect their anonymity, which makes data collection problematic. Assessing the scale of informal activities calls for great accuracy. Moreover, informal employment can take many forms: there are workers who have an informal job in addition to being formally employed; there are those who choose to work "off the books"; and, finally, workers who are not allowed to work legally (such as immigrants) \citep{Schneider2002}. A simple division into the formal and informal sector is~impossible, since there are many forms of informal work which occurs in registered enetrprises. According to the definition proposed by the Organisation for Economic Co-operation and Development (OECD), informal employment covers \textit{employment engaged in the production of legal goods and services but where one or more of the legal requirements usually associated with employment are not met} \citep{Venn2008}. According on the above definition, informal workers include those who are not registered for mandatory social security, who are paid less than the legal minimum wage or work without a~written contract, falsely  self-employed, those underreporting their income and entirely unregistered companies \citep{Venn2008}. At the same time, according to estimates of Statistics Poland, the undeground economy accounted for 12.9\% of Poland's GDP in 2016,with 10.8\% of GDP generated by unreported economic activity in registered entities \citep{StatisticsPoland2018}.

Similarly, the survey module „Nonstandard forms of employment and unregistered employment” carried out in parallel with the Polish Labour Force Survey (LFS) in 2014 indicates that 26.5\% of employers offering unregistered work were private companies \citep{CSO2015}. Importantly, most available data on informal employment come from surveys, which implies an underestimated level of the phenomenon, because respondents are not willing to report their informal activities. It has been assumed that survey-based studies of informal employment and the shadow economy indicate only the lower boundary of this phenomenon. Moreover, other disadvantages of survey data include high non-response rates, probably owing to the topic of the survey, and dishonest responses \citep{ILO2013}. The informal economy is~also studied from the enterprise perspective, but such studies are much less frequent. The fifth round (2012-2016) of The Business Environment and Enterprise Performance Survey (BEEPS), which is~a~firm-level survey of a~representative sample of the private sector, shows that almost 27\% of surveyed enterprises in Poland compete against unregistered or informal firms\footnote{Data available on \url{https://ebrd-beeps.com/data/2012-2016/}}. 


In view of the above, there is~a~need for an alternative measure of informal activities. In this study we propose using enterprise administrative data on informal employment in order to estimate the prevalence of this phenomenon. We used data from two administrative sources: the National Labour Inspectorate (later on referred to by its Polish acronym: PIP) and The Polish Social Insurance Institution (later on referred to by its Polish acronym: ZUS) to assess the magnitude of informal work conducted in registered  enterprises in Poland. In 2016 PIP inspected 26,261 companies employing a~total of 161,927 workers, of whom 21,285 were employed illegally on the day of or prior to the inspection. The illegal status is~defined as employment without a~formal agreement or lack of registration for mandatory social security. This makes PIP data the biggest source of information about the grey economy in Poland. Despite the size, the source has two main shortcomings. 

First, data collected by PIP are a~purposive sample, i.e. one that includes companies that are more likely to employ informal workers. Moreover, PIP is~divided into 16 District Labour Inspectorates, which independently inspect companies and their staff (for more information see \ref{app-nli-det}).  As a~result, these data are highly non-representative of the whole economy. To overcome this limitation, we view the problem in the context of the missing data mechanism and treat the selection mechanism as non-ignorable \citep{rubin1976inference}. In our study we applied a~bivariate probit-logit sample selection model assuming non-Gaussian (copula) correlation between errors resulting from selection and the outcome model. This approach is~an extension of Heckman's classical sample selection model \citep{heckman1979, marra2017joint, wojtys2016copula}.

Second, definitions of the target variable and the population are based on the act regulating the functioning of PIP and other acts regarding the organization of the labour market. PIP inspectors have the right to inspect documents on employment up to three years back. Unfortunately, the reporting form used by inspectors does not contain such information, so it is~not possible to identify the reference period. Fortunately, PIP categorizes employees into two groups: current and former employees, with current employees further divided into illegally employed on the day of or prior to the inspection. PIP collects data separately for Polish citizens and foreigners. Our study was limited to inspections concerning Polish citizens who were currently employed.

We defined the the target population as enterprises that are active payers of health insurance contributions according to the ZUS register. For more details regarding the definition, see Appendix \ref{payer-def}. The reason for choosing this definition of the target population is~given later on.

The article provides two contributions to the literature. First, we propose an approach to estimating informal employment, which is~based solely on register data in order to increase the reliability of the results. Second, we focus on the population of firms and analyse company characteristics (such as size) that increase the probability of using informal employees. We then apply a~novel technique that accounts for a~non-Gaussian correlation of errors in the selection and the outcome equation and clustering by incorporating random effects in a~generalization of Heckman's sample selection model.

We found that 5.7\% (4.6\%,7.1\%; 95\% CI) of registered enterprises in Poland to some extent take advantage of informal labour force. Our study exemplifies a~new approach to measuring informal  employment,  which  can  be  implemented  in  other  countries. It also contributes to the existing literature by providing, to the best of our knowledge, the first estimates of informal employment at the level of companies based solely on administrative data.

The structure of this article is~as follows: section \ref{lit-ref} presents a~review of the literature on informal work in Poland, section \ref{admin-data} describes administrative data, section \ref{methods} describes approaches to dealing with the non-ignorable selection mechanism and the selected method, section \ref{results} shows our results and section \ref{summary} is~the conclusion. 


\section{Review of literature on informal work}\label{lit-ref} 

First of all, it is~necessary to draw a~distinction between the way the informal sector is~measured in developed and developing countries, which is~connected with different determinants of informal activities in each group of countries \citep{Goel2016}. The informal sector in low and medium-income countries accounts for ca. 40\% of non-agriculture employment, or an even higher percentage in Sub-Saharan Africa, Latin America and the Caribbean and South and East Asian countries \citep{ILO2011}. This phenomenon is~mainly associated with unregistered and unregulated small-scale activities conducted by the urban poor \citep{Bernabe2002}. Therefore, the majority of available literature is~devoted to informal employment in developing countries. In developed countries the informal sector is~considerably smaller but more heterogeneous and less documented. In view of the above, our review is~limited to the existing studies concerning developed countries, mostly European ones, given the fact that our novel approach of measuring informal employment was applied to the Polish economy. 
 
Despite the clandestine nature of informal activities, survey estimates are still a~popular way of measuring the scale of informal employment. Since the main focus of our study was to measure informal employment, not the informal (shadow) economy  as a~whole, below we mention the exiting methods of measuring informal employment. A cross-country labour force survey of undeclared work in the European Union was conducted by the European Commission in 2007 and 2013 \citep{EuropeanCommission2007, EuropeanCommission2014}. In the newest release (2014) 4\% of respondents reported carrying out undeclared paid activities apart from regular employment in the past year \citep{EuropeanCommission2014}. Most studies are based on national surveys using proxies for informal employment. \citet{Bernabe2002} used data from the Georgia Labour Force Survey (1998, 1999) to distinguish the following categories of informal workers: own-account workers and employees working in household enterprises, unpaid contributing family workers, workers employed under oral agreements, workers employed casually or temporarily and workers with formal primary jobs and informal secondary jobs. \citet{DiCaro2016} measured informal employment using data from the Istituto Nazionale di Statistica (ISTAT) for 20 Italian regions over the period 2001–12. They found significant differences in the level of underground employment between Italian regions: from 7\% in Lombardy to about 21.5\% in Sicily. \citet{Lehmann2015} used results from the Russian Longitudinal Monitoring Survey (RLMS) and defined informal employees as those not officially registered.

Another body of research is~based on surveys dedicated to the specific issue of informal employment or the shadow economy in general. It should be noted that not much research of this kind has been conducted. \citet{Pedersen2003} carried out a~survey in Denmark, Norway, Sweden, Germany and Great Britain and found that the share of "off the books" hours worked to formal working hours ranges from 8.7\% in Great Britain to 21.6\% in Denmark. \citet{Merikull2010} examined the prevalence and determinants of unreported employment in three Baltic countries in 1998 and 2002 using several questions related to "salary in envelope” or “black salary”. They found that firm-related characteristics, such as type of activity, firm size and employment changes, have a~considerable influence on the prevalence of unreported employment. \citet{Kriz2008} compiled data on informal employment from three different sources: a~survey of self-reported tax evasion by the Estonian Institute for Economic Research, compliance data from audits of individuals by the Estonian Tax and Customs Board and The Labour Force Survey conducted by Statistics Estonia. Their findings indicate that the magnitude of informal employment differs significantly depending on the source (2.7\% of respondents reporting not having a~written contract, 66\% of audited individuals were found to evade taxes and 14.4\% of surveyed employees reported being paid under the table). 

In one of more recent studies  \citet{Petreski2018} analyses the relationship between informal employment of youth and their later labor market outcomes using School to Work Transition Surveys of the ILO. Based on data from transition economies of Southeast Europe (SEE) and the Commonwealth of Independent States (CIS) he found that early experience of informal work negatively affected further employment prospects. \citet{Krasniqi2017} used 2010 Life in Transition Survey (LiTS) involving 35 Eurasian countries to study informal employment defined as employment without a~written contract. They reported that countries with a~higher tax morale, higher GDP per capita and a~higher level of social distribution and state intervention experience lower levels of informal employment. \citet{Florez2016} used data from the European Social Survey for over 30 European countries to examine informal employment defined as employment without a~formal contract. He found that in countries of Southern and Western Europe employment protection and unemployment benefits reduce the prevalence of informal work. \citet{Besim2015, Ekici2018} used household survey and census data to estimate the number of informal workers in Northern Cyprus. They defined informal workers as those who are not officially registered with any social security scheme.

Because of the obvious challenges in measuring informal activities, some researchers use a~novel alternative approach to track the informal sector. \citet{Autio2015} investigated the impact of economic and political institutions on the prevalence rate of formal and informal entrepreneurship across 18 countries in the Asia-Pacific region during the period 2001–2010 and calculated a~informal businesses entry rate as a~difference between Global Entrepreneurship Monitor (GEM)-based estimate of all new entries and the World Bank Enterprise Snapshot estimate of new business. 

Customs Administration and the Central Administration of National Pension Insurance. \citet{Porto2011} used individual audit data from the labor inspectorate to estimate the undeclared labor force in the Piedmont region in Italy.

There are few estimates of the size of informal employment in Poland because of the scarcity of national statistical data. According to annual estimates published by Statistics Poland, 2.1\% of the country's GDP is~generated by individuals who performed non-registered work, mainly in services, including persons providing prostitution services \citep{StatisticsPoland2018}. Results from the survey of „Nonstandard forms of employment and unregistered employment” conducted alongside the Labour Force Survey (LFS) indicate that 4.5\% of the total number of employees are informal workers without a~written employment contract \citep{CSO2015}. According to another survey -- the Human Capital Balance -- 7.4\% of economically active people reported working under a~non-formal agreement \citep{Balance2017}. Apart from national sources, there are some international surveys designed to measure informal employment in Europe. Accrding to the Eurobarometer survey 2013 conducted by Eurostat, 3\% of employees in Poland were working without a~formal contract \citep{EuropeanCommission2014}. The European Social Survey data (2006/2007 wave) indicates that 5.9\% of all employees in Poland can be treated as informal, while the Fourth European Working Conditions Survey (EWCS 2005) estimates the share of informal employment at the level of 6.5\%  \citep{Hazans2011}. The above review shows that measurements of informal employment based on survey data yield different results and therefore may be less reliable than estimates based on administrative data. 

\section{Data}\label{admin-data}

To estimate the prevalence of informal employment we used administrative data from inspections conducted by the National Labour Inspectorate (PIP). PIP is~an authority established to supervise and enforce compliance with the Labour Code. A detailed description of the PIP structure its data collection process is~provided in Appendix \ref{source-desc}. As mentioned in the introduction, companies to be audited are not selected at random, so this mechanism must be taken into account. According to PIP authorities, companies are selected taking into account the following \citep{ChiefLaborInspectorate2017}:

\begin{itemize}
    \item industries with the largest number of irregularities detected in previous years,
    \item entities employing a~small number of workers (micro and small enterprises),
    \item complaints and requests sent to PIP,
    \item notifications from other authorities,
    \item the monitoring of the media (e.g. online forums),
    \item knowledge of local conditions and the specifics of the labor market,
    \item analyses obtained from units of local government.
\end{itemize}

During the selection procedure PIP uses basic information from the National Official Business Register (later on referred to by its Polish acronym: REGON) maintained by Statistics Poland. The variables include address, NACE code (for the main type of activity) and company size. 

The REGON register is~not suitable as a~source for population data because of the following drawbacks. Once a~company is~included in the register, its information is~rarely updated. This leads to overcoverage resulting from the lack of up-to-date information about the company's activity and a~measurement error in the number of employees relative to what is~declared at the time of registration. These problems concern mainly micro and small firms.

Therefore, we decided to use data from the Social Insurance Institution (ZUS) to describe the target population. ZUS is~a~state organizational unit with its own legal status , which collects citizens' social insurance and health contributions and distributes benefits and allowances (e.g. old-age pensions, disability pensions, sickness benefits or maternity benefits) and maintains individual insurance accounts.

From ZUS we obtained population-level (aggregated) data about active payers of health insurance contributions classified by industry (NACE code), number of insured persons at district level as at 31.12.2016. Assignment to province and district according to the registered head office of the payer is~consistent with PIP data. The NACE code comes from the REGON register. More details about these two sources can be found in Appendix~\ref{source-desc}.

Thanks to the co-operation with PIP we obtained aggregated data from 2010 to 2016 and anonymized unit-level data for 2016 containing information about companies' NACE codes (\textit{the Statistical Classification of Economic Activities in the European Community} - NACE\footnote{fr. \textit{Nomenclature statistique des activités économiques dans la Communauté européenne}}) and company size measured by the number of employees (4 groups; up to 9 at district level (380 administrative units grouped into 16 provinces).

Finally, because of a~small number of inspections in 2016 we decided to exclude from the population the following industries: Mining and quarrying (Section B); electricity, gas, steam and air conditioning supply (Section D) and public administration and defence, compulsory social security (Section O). We also excluded contribution payers that only paid their insurance contributions. The data preparation procedure is~described in Appendix \ref{data-prep}.

PIP registers show that in 2016 there were almost 26,000 inspections, which accounted for almost 3\% of the population consisting of around 870,000 companies (according to ZUS). Na\"{\i}ve estimate of informal employment according to PIP data was almost 32\%.

\begin{figure}[ht!]
    \centering
    \includegraphics[width=0.6\textwidth]{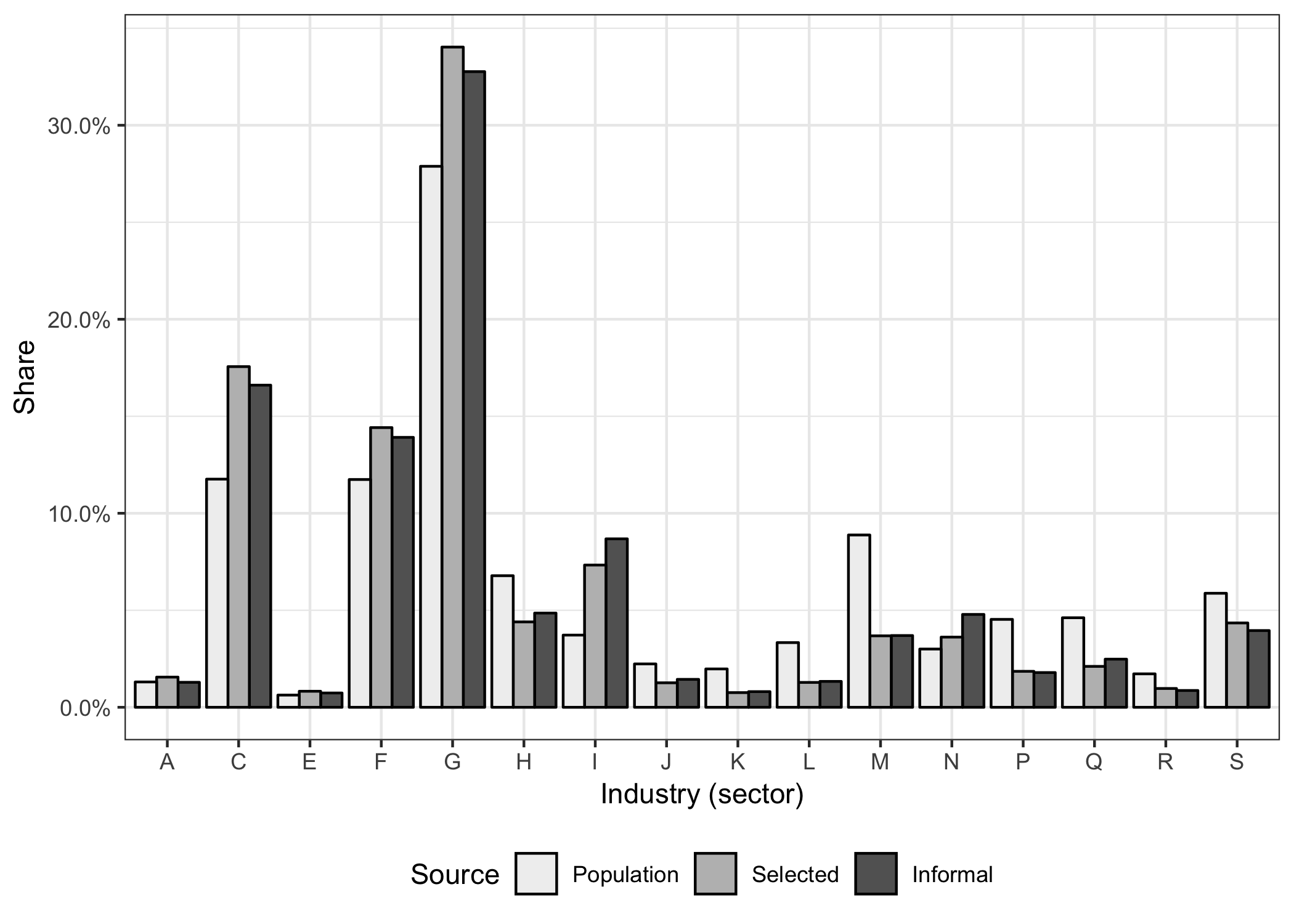}
    \includegraphics[width=0.6\textwidth]{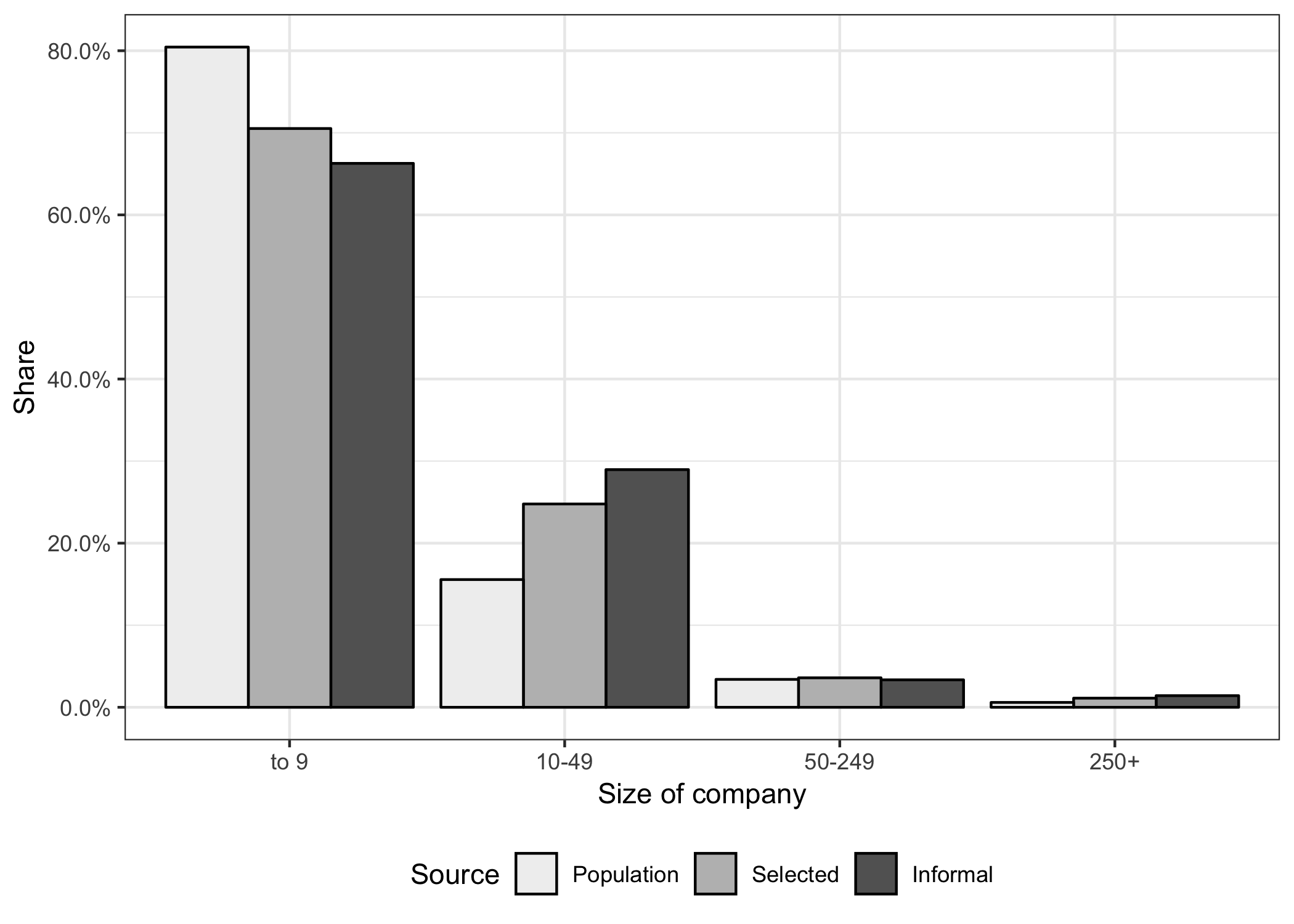}
    \caption{Prevalence of informal workers based on PIP data by company size and NACE section in 2016. Description of sections are available in Appendix \ref{app-nli-det}}
    \label{comp-size}
    \begin{flushleft}
\end{flushleft}
\end{figure}

Figure \ref{comp-size} presents discrepancies between presents discrepancies between the share that each section accounted for within the target population, the group of companies selected by PIP and those which employed illegally. Within each group percentages add up to 100. Detailed information is~presented in the Appendix \ref{app-nli-det}. 

The biggest number of inspections were held in the following sections: wholesale and retail trade, repair of motor vehicles and motorcycles  (34\%, section G), manufacturing  (18\%, section C) and construction (14\%, section F) since those sections are believed to be the ones most affected by informal employment. Indeed, the shares of informal workers inspected in those sections are the highest in the whole economy: 33\% in section G, 17\% in section C and 14\% in section F, which confirms the initial assumptions about the prevalence of informal workers in those sections. This means there is~a~correlation between characteristics of companies that are selected and those that do not comply with the regulations.

Moreover, special attention is~paid to companies employing a~small number of employees (micro and small enterprises), which are the ones most frequently found in violation of the regulations regarding the legality of employment.  In 2016 most inspections were held in companies employing up to 9 people (70 \%) and from 10 to 49 (24\%). Firms in the second and fourth group are oversampled in the PIP selection. In fact, the magnitude of informal employment can mainly be attributed to micro enterprises, where informal workers account for over 66\% of all employees, and in small enterprises where the share of informal workers is~almost 30\%. As expected, the prevalence of informal employment is~mainly associated with micro and small enterprises, while the phenomenon is~rather negligible in medium and large companies..

While the share of companies selected for inspections and found to be non-complying across sections was similar, there are differences when company size is~taken into account. The share of companies in the second group (10-49 employees) and fourth group (250+) found to employ illegally was actually greater than the percentage selected for inspection, while the opposite was true for micro companies.

\section{Estimation under a~non-ignorable selection mechanism}\label{methods}

There are several approaches to dealing with non-random samples with a~non-ignorable selection mechanism (i.e. not missing at random; NMAR). \citet{Sikov2018} distinguishes three groups of techniques: 1) \textit{weighting procedures} that adjust for non-response by assigning weights to responding units; 2) \textit{imputation-based procedures} to create a~complete data set by replacing missing values with plausible values and 3) \textit{model-based procedures}, where the inference is~carried out by using models fitted to the observed data. For recent reviews and research see \citet{Sikov2018, Tang2018, kim-ying-2018}. 

The problem is~not new and first attempts to tackle it were made in late 1950s by James Tobin and then in 1970s by James Heckman. \citet{heckman1979} proposed a~model that can account for the selection mechanism by jointly modelling the selection and target variable assuming a~Gaussian correlation of errors between these equations. That model was further developed to account for other distributions of errors, notably copulas, and assuming semi- and non-parametric models.

Another approach was taken by \citet{sverchkov2008new, Pfeffermann2011, riddles2016propensity}, who considered a~model for the outcome variable assuming complete response and a~model for the response probability, which is~allowed to depend on the outcome and auxiliary variables. When combined, the two models yield a~"respondents model".  In all of these approaches it is~important to have an~instrumental variable that is~associated either with the selection or the target variable but not with both (see \cite{riddles2016propensity, Wang2014a}).

In the article we also discuss an extension of Heckman's sample selection model, which takes into account a~non-Gaussian distributions of errors, a~family of exponential distributions and random effects. There are examples in the literature on official statistics that take into account such models to correct the selection error (for example, copulas are discussed in \citet{dalla2016use}) but as far as we know the use of these models is~limited. 


\subsection{Generalized Heckman-type sample selection models}


\subsubsection{Specification of the model}

In this section we follow the notation introduced by \citet{marra2017joint}. Let $(Y_{i1},Y_{i2})$, for $i = 1,..., N$, where $N$ represents the size of the target population. The probability of event $(Y_{i1}=1, Y_{i2} = 1)$ can be defined as:

    \begin{equation}
p_{11 i}=P\left(Y_{i 1}=1, Y_{i 2}=1\right)=C\left(P\left(Y_{i 1}=1\right), P\left(Y_{i 2}=1\right) ; \theta\right),
\end{equation}

\noindent where $P(Y_{ij}=1) = 1 - F_j(-\eta_{ij})$, for $j=1,2$ and $F(\cdot)$ is~the cumulative distribution function (cdf) of a~standardized univariate distribution (e.g. binomial), $\eta_{ij} \in \mathbb{R}$ is~an additive predictor, $C$ denotes a~bivariate copula and $\theta_i$ is~an association parameter measuring the dependence between the two random variables.

Further, let $Y^*_{ij}$ be a~continuous latent variable associated with $P(Y_{ij}=1)$, which is~given by:

\begin{equation}
    Y^*_{ij} = \eta_{ij} + \epsilon_i, 
\end{equation}

\noindent where $\epsilon_i$ is~an error term and $Y_{ij}$ can be treated as an indicator variable $Y_{ij}^* >0$. Therefore, $P(Y_{ij}=1) = P(Y^*_{ij} >0) = 1 - F_j(-\eta_{ij})$.

The log-likelihood function of the sample can be expressed as:

\begin{equation}
\ell=\sum_{i=1}^{n}\left\{\sum_{a, b=0}^{1} \mathbb{1}_{a b i} \log \left(p_{a b i}\right)\right\},
\end{equation}

\noindent where $\mathbb{1}_{a b i}$ is~an indicator function equal to 1 when $(y_{i1}=a,y_{i2}=b)$ is~true, $a,b \in \{ 0, 1\} $ and $y_{i1}, y_{i2}$ are realisations of $Y_{i1},Y_{i2}$ respectively.

We assume that $Y_1$ and $Y_2$ follow the Binomial distribution and are given by the following set of equations:

\begin{equation}
    Y_{i1} = \begin{cases}
    1 & \text{if company $i$ is~selected by PIP}, \\
    0 & \text{otherwise},
    \end{cases}
\end{equation}

\noindent and 

\begin{equation}
    Y_{i2} = \begin{cases}
    1 & \text{if company $i$ employs at least one person informally},\\
    0 & \text{otherwise}.
    \end{cases}
\end{equation}

The selection and outcome equation can be modelled with  \textit{probit}, \textit{logit} or \textit{cloglog} link function. The question of which of these functions to select will be discussed in section \ref{results}.

\subsubsection{Specification of the additive term}

Let $\eta_{ji}$ be a~generic additive predictor and overall vector of covariates denoted by $\bz_{ji}$. The predictor can be expressed as:

\begin{equation}
\eta_{j i}=\beta_{j0}+\sum_{k_{j}=1}^{K_{j}} s_{j k_{j}}\left(\bz_{j k_{j} i}\right), i=1, \ldots, N,
\label{eq-additive-term}
\end{equation}

\noindent where $\beta_{j0} \in \mathbb{R}$ is~an overall intercept, $\bz_{j k_{j} i}$ denotes the $k^{th}_j$ sub-vector of the complete vector of $\bz_{ji}$ and the $K_j$ functions $s_{j k_{j}}(\bz_{j k_{j} i})$ represent generic effects which are chosen according to the type of covariate(s) considered. Each $s_{j k_{j}}(\bz_{j k_{j} i})$ can be approximated as a~linear combination of $Q_{jk_j}$ basic functions $b_{jk_jq_{jk_j}}(\bz_{ji})$ and regression coefficients $\beta_{jk_jq_{jk_j}} \in\mathbb{R}$, namely:

\begin{equation}
    s_{j k_{j}}(\bz_{j k_{j} i}) = \sum_{q_{jk_j}=1}^{Q_{jk_j}} \beta_{jk_jq_{jk_j}} b_{jk_jq_{jk_j}}(\bz_{jk_ji}).
\end{equation}
 
\noindent The notation can be simplified in the following way. Let $\bZ_{jk_j}\bbeta_{jkj}$ be defined as:

\begin{equation}
\bZ_{jk_j}\bbeta_{jkj} = \left\{s_{j k_{j}}\left(\bz_{j k_{j} 1}\right), \ldots, s_{j k_{j}}\left(\bz_{j k_{j} N}\right)\right\}^{\top},
\end{equation}

\noindent with $\boldsymbol{\beta}_{j k_{j}}=\left(\beta_{j k_{j} 1}, \ldots, \beta_{j k_{j}Q_{j k_{j}}}\right)^{\top}$ and design matrix $\bZ_{jk_j}[i, q_{jk_j}] = b_{jk_jq_{jk_j}}(\bz_{jk_ji})$.  Given this, we can rewrite \eqref{eq-additive-term} as:

\begin{equation}
\beeta_{j}=\beta_{j 0} \mathbf{1}_{N}+\bZ_{j 1} \bbeta_{j 1}+\ldots+\bZ_{j K_{j}} \bbeta_{j K_{j}} = \bZ_j \bbeta_J,
\end{equation}

\noindent where $\mathbf{1}_{N}$ is~an $N$-dimensional vector made up of ones. 

In our calculations we used the implementation of this model in the \texttt{GJRM} package \citep{marra2017joint} and the \texttt{SemiParSampleSel} \citep{wojtys2016copula} in the R language  \citep{rcran}. The packages can be used to specify linear and random effects as well as non-linear effects. The implementation makes use of the \texttt{mgcv} package (for more details see \cite{wood2017}).

\subsubsection{Specification of Copulas}\label{copulas}

\begin{table}[ht!]
\centering
\small
\caption{Definitions of copulas used in the modelling procedure}
\label{tab-copulas}
\begin{tabular}{lccc}
\hline
Copula  & $C(u, v; \theta)$ & $\theta$ space & Kendall's $\tau$\\
\hline
Normal  & $\Phi_{2}\left(\Phi^{-1}(u), \Phi^{-1}(v) ; \theta\right)$ &  $\theta \in [ -1,1]$ 
& 
$
\frac{2}{\pi} \arcsin \left(\theta\right)
$
\\
Clayton & $\left(u^{-\theta}+v^{-\theta}-1\right)^{-1 / \theta}$ & $\theta \in (0, \infty)$
& 
$
\frac{\theta}{\theta+2}
$
\\          
Joe & $1-\left[(1-u)^{\theta}+(1-v)^{\theta}-(1-u)^{\theta}(1-v)^{\theta}\right]^{1 / \theta}$ & $\theta \in (1, \infty)$ 
& 
$
1+\frac{4}{\theta^{2}} D_{2}\left(\theta\right)
$
\\         
Gumbel    &  $\exp \left\{-\left[(-\log u)^{\theta}+(-\log v)^{\theta}\right]^{1 / \theta}\right\}$ &  $\theta \in [1, \infty)$ 
& 
$
1-\frac{1}{\theta}
$
\\  
AMH    & $u v /[1-\theta(1-u)(1-v)]$ & $\theta \in [ -1,1]$
& 
$
\begin{array}{l}{-\frac{2}{3 \theta^{2}}\left\{\theta+\left(1-\theta\right)^{2}\right.} \\ {\log \left(1-\theta\right) \}+1}\end{array}
$\\  
\hline
\end{tabular}
\begin{flushleft}
\label{copulas-table}
\textit{Note:} Families of copulas implemented in \texttt{SemiParSampleSel}, with the corresponding parameter range of association parameter $\theta$. $\Phi_{2}(\cdot, \cdot; \theta)$ denotes the cumulative distribution function of a~standard bivariate normal distribution with correlation coefficient $\theta$ and $\Phi(\cdot)$ denotes the cdf of a~univariate standard normal distribution. Kendall's $\tau$ includes $D_{2}\left(\theta\right)=\int_{0}^{1} t \log (t)(1-t)^{\frac{2\left(1-\theta\right)}{\theta}} dt$.
\end{flushleft}
\end{table}

Generalized Heckman's sample selection model assumes that the relationship between the selection and outcome equations is~described by copulas denoted by $C(\cdot, \cdot; \theta)$. The appropriate relationship should be selected in line with expectations about the selection process. Since we expected a~positive correlation between $Y_1^*$ and $Y^*_2$ because of the design of the purposive sample, we selected copulas that assume this type of relationship, namely: Clayton, Gumbel, Joe and Ali-Mikhail-Haq (AMH) copulas. As a~reference, we used standard Normal copula that assumes the bivariate normal distribution. For this reason, our description of copulas is~limited to the these five. 

Table \ref{copulas-table} provides definitions of the five copulas including their distribution, parameter space and Kendall's $\tau$. Only normal and AMH copulas have the parameter space ranging from -1 to 1, which may indicate a~negative or positive relationship between two distributions. In order to verify the correlation between the selection and outcome variables for all the copulas, the Kendall's $\tau$ measure should be used (cf. \cite{wojtys2018copula}).

\subsection{Estimators}

In our study we were interested in estimating the prevalence of informal work in companies in Poland, so the target quantity is~given by:

\begin{equation}
    \overline{Y}_{\text{prev}} = \frac{1}{N} \sum_{i=1}^N Y_2,
\end{equation}

\noindent where $N$ denotes sthe ize of the target population and $Y_2$ is~defined as previously. As we did not observe $Y_2$ for the whole population, $\overline{Y}_{\text{prev}}$ should be estimated based on PIP data. We compared the proposed estimator based on a~generalized Heckman's sample selection model with estimators assuming an ignorable selection mechanism. 

\begin{itemize}
    \item Proposed estimator (generalized Heckman's; GH) is~given by \eqref{eq-estim-ghss}
    
    \begin{equation}
        \widehat{\overline{Y}}_{\text{prev}}^{\text{GH}} =  \frac{1}{N} \sum_{i=1}^N (1 - F_2(\hat{\eta}_{2i}))
        \label{eq-estim-ghss}
    \end{equation}
    where $1 - F_2(\hat{\eta}_{2i})$ is~the estimated prevalence of informal work based on the generalized Heckman's model for each unit $i$ in the target population. We also report results for specific domains denoted by $d$ of size $N_d$ and the estimator is~given by:
    
    \begin{equation}
        \widehat{\overline{Y}}_{\text{prev},d}^{\text{GH}} =  \frac{1}{N_d} \sum_{i=1}^{N_d} (1 - F_2(\hat{\eta}_{2i})).
        \label{eq-estim-ghss-domain}
    \end{equation}
    
    \item The na\"ive, i.e. simple average, based on PIP data is~given by the following formula:
    \begin{equation}
        \widehat{\overline{Y}}_{\text{prev}}^{\text{Na\"ive}} = \frac{1}{N} \sum_{i=1}^{N} Y_{1i}Y_{2i}.
    \end{equation}
    \item Propensity-score weighted estimator is~given by the following formula:
    \begin{equation}
        \widehat{\overline{Y}}_{\text{prev}}^{\text{PS}} = \frac{1}{N} \sum_{i=1}^{N} \frac{Y_{1i}Y_{2i}}{\pi\left(Y_{1i} = 1 |\boldsymbol{x}_{i}; \hat{\boldsymbol{\phi}}\right)},
    \end{equation}
    where $\pi(\cdot; \hat{\boldsymbol{\phi}})$ denotes the propensity score of being observed in the PIP sample under an ignorable selection mechanism and $\hat{\boldsymbol{\phi}}$ is~a~vector of parameters of the propensity score model. We used logistic regression to estimate $\pi(\cdot)$ .
\end{itemize}

We provide variance estimates only for the proposed estimator \eqref{eq-estim-ghss} using a~Bayesian posterior simulation described in \citet[section 2.3]{marra2017joint}. 

The process of model selection was based on AIC given by $-2\ell(\hat{\bdelta})+2edf$ and BIC $-2\ell(\hat{\bdelta})+\log(N)edf$, where $\ell$ is~evaluated using penalized parameter estimates, $edf$ represents effective degrees of freedom and $\bdelta = (\bbeta_1,\bbeta_2, \theta)$ is~a~parameter vector estimate including the additive term and copula parameters.

\section{Results}\label{results}

\subsection{Description of the models}

Prior to choosing appropriate copulas for the selection model to estimate the prevalence of informal work we selected variables for the selection and the outcome equation.  \texttt{district} (380 Local Administrative Units Level 1; LAU 1) and \texttt{complaints} (the number of complaints sent to PIP relative to the total number of companies at district level) were used as instrumental variables associated with the selection mechanism while \texttt{unemployment} (i.e. the unemployment rate at district level) was used as an additive term associated with the outcome variable with thin plate splines. NACE section and company size, as defined in \ref{app-nli-det}, were used in both models.

The same set of variables was chosen for the propensity weighted estimator to estimate the probability of selection using a~generalized linear mixed model. 

First, we determined which link functions should be used to model the selection and outcome variables. We calculated 20 models, obtained by combining 4 link functions (\textit{logit}, \textit{probit}) and the five copulas. As can be seen in Table \ref{tab-link-funs}, the lowest AIC values were obtained for the \textit{probit} link function in the case of the selection and for the \textit{logit} link function with respect to the outcome equation. We chose this model in the subsequent stage of the estimation process.

Then, using the selected variables, we compared five models to choose appropriate copulas from the ones discussed in section \ref{copulas}. The selection was based on AIC and BIC and the results are shown in Table \ref{tab-model-selection}. The table also includes information about the estimated prevalence of informal employment ($\widehat{\overline{Y}}_{\text{prev}}^{\text{GH}}$). Based on these results we selected the Gumbel copula because of the lowest value of the information criteria. The estimated prevalence from this model was 5.75\%.

\begin{table}[ht!]
\centering
\caption{Model selection based on information criteria for the probit-logit model}
\label{tab-model-selection}
\begin{tabular}{rlrrr}
  \hline
 Copula & AIC & BIC & $\widehat{\overline{Y}}_{\text{prev}}^{\text{GH}}$  \\
  \hline
  Gumbel & 249,473 & 253,035 & 5.67 \\
   Ali-Mikhail-Haq & 249,475   & 253,039 & 4.74 \\
    Normal & 249,481 & 253,042 & 7.36 \\
    Joe & 249,483 & 253,046 & 3.10 \\
    Clayton & 249,483 & 253,044 & 11.54 \\
   \hline
\end{tabular}
\end{table}

\begin{figure}[ht!]
\centering
\includegraphics[width=0.32\textwidth]{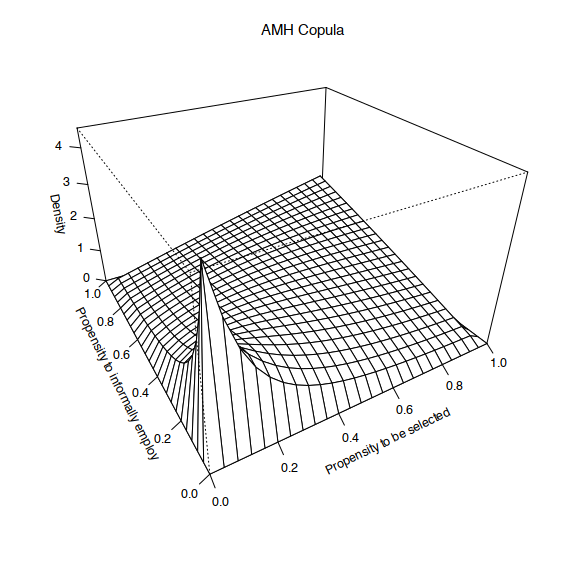}
\includegraphics[width=0.32\textwidth]{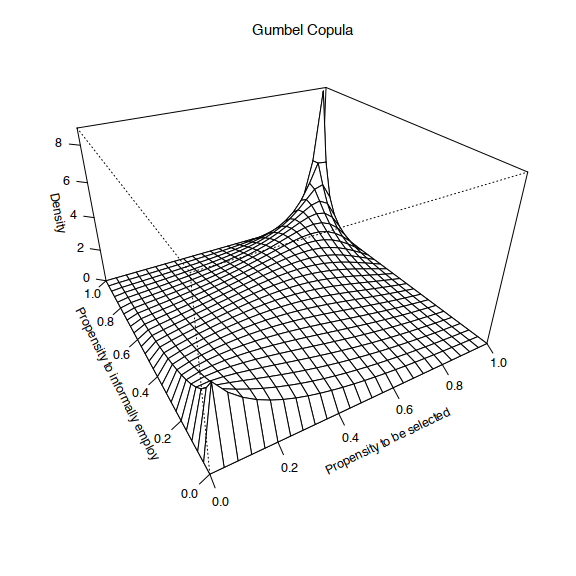}
\includegraphics[width=0.32\textwidth]{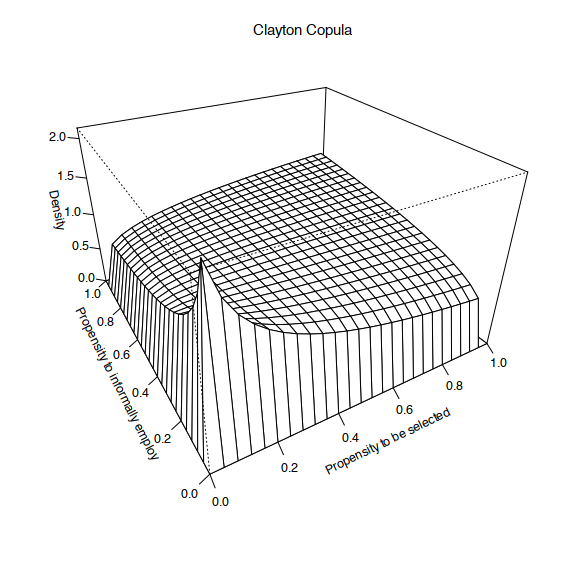}
\includegraphics[width=0.32\textwidth]{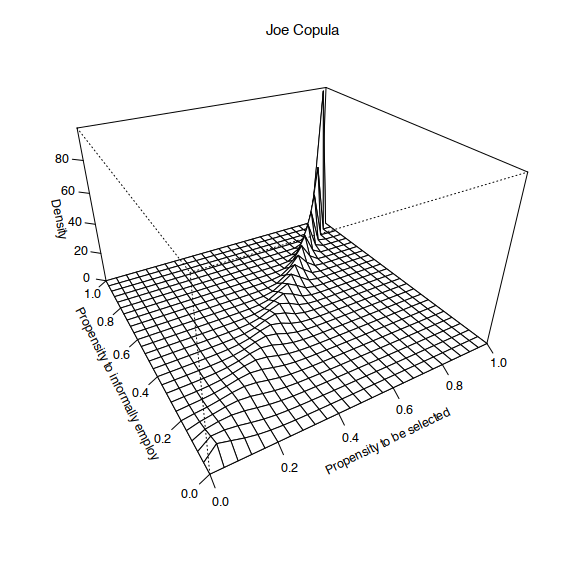}
\includegraphics[width=0.32\textwidth]{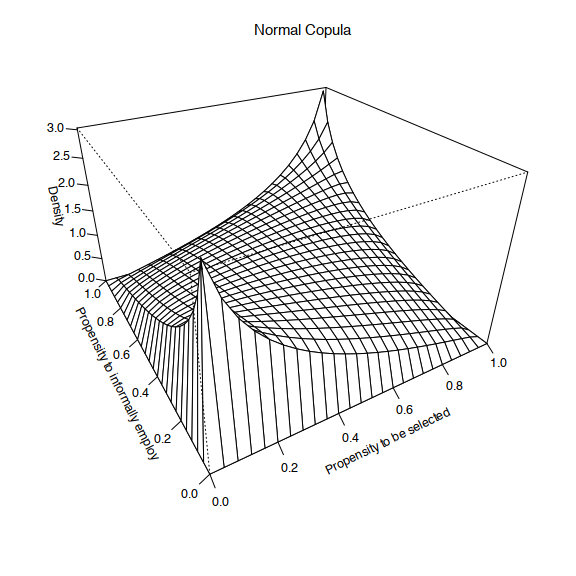}
\caption{Copulas obtained from the models used in the study. The following parameters were estimated for these copulas: AMH($\theta$=0.934), Clayton($\theta$=0.23), Joe($\theta$=16.1), Gumbel($\theta$=1.93) and Normal($\theta$=0.472)}
\label{fig-copulas-vis}
\end{figure}

We also compared the models by visual inspection of the copulas and the results in Figure \ref{fig-copulas-vis}. The AMH and Clayton copulas suggest that companies not selected for inspections are more likely not to employ illegally, which is~indicated by the concentration of the bivariate distribution near the $(Y_1=0,Y_2=0)$ point.  This is~in line with the underlying selection mechanism used by PIP to select companies for verification.

The Gumbel and Joe copulas indicate that companies that have a~higher prevalence of informal employment are more likely to be audited by PIP. The Joe copula shows a~higher correlation between the selection and outcome equations but is~characterised by the highest AIC and BIC. 

Finally, as expected, the Normal copula suggests that there are no differences between companies selected for inspection in terms of whether they employ formally and informally. Both ends of the bivariate distribution are equally likely.

Keeping that in mind, we chose a~generalized Heckman's sample selection model with the Gumbel copula. A detailed description of the estimated model is~presented in Appendix \ref{model-details}. We used this model for further comparisons and for the estimation of the  prevalence of informal employment in Poland.  Table \ref{tab-estimators-results} contains a~comparison of point estimates of the model and two estimators assuming an ignorable selection mechanism. 

\begin{table}[ht!]
    \centering
    \caption{Comparison of estimators of the prevalence of informal work in Poland in 2016}
    \label{tab-estimators-results}
    \begin{tabular}{lrrr}
    \hline
    Estimator & 
        $\widehat{\overline{Y}}_{\text{prev}}^{\text{Na\"{\i}ve}}$ & 
        $\widehat{\overline{Y}}_{\text{prev}}^{\text{PS}}$ &  
        $\widehat{\overline{Y}}_{\text{prev}}^{\text{GH}}$\\
    \hline
    Point estimate & 31.99 & 34.54 & 5.67\\
    \hline
    \end{tabular}
\end{table}

According to the na\"{\i}ve estimator, the prevalence of informal work was equal to 32\%, which means that a~third of companies in Poland engaged in illegal employment - an obvious overestimation. The propensity score weighted estimator is~equal to 34.54\%, which is~even higher than the na\"{\i}ve estimator. The estimate of nearly 6\%, produced by the generalized Heckman's model, seems to be more plausible and means that PIP correctly selected 14.5\% of such companies. A detailed comparison of the generalized selection model is~presented in Table \ref{tab-models}.  

\subsection{Prevalence of informal work}

Based on the GH model we estimated the magnitude of informal employment in Polish registered enterprises. Our results show that 5.67\% (4.59,7.10) of companies active in 2016 employed at least 1 person off the books (Table \ref{tab-estimates-results}).

\begin{table}[ht!]
    \centering
    \caption{Estimates of total prevalence of informal employment and broken down by company size in 2016 based on the sample selection (probit-logit) model with the Gumbel copula}
    \label{tab-estimates-results}
    \begin{tabular}{lrrr}
    \hline
    Classification & Estimate & 5\% &  95\% \\
    \hline
    Total & 5.67 & 4.59 & 7.10\\
    \hline
    to 9 & 5.12 & 3.96 & 6.49 \\ 
    10-49 & 8.49 & 6.73 & 10.54 \\ 
    50-250 & 5.12 & 3.93 & 6.85 \\ 
    250+ & 9.27 & 6.92 & 12.49\\ 
    \hline
    \end{tabular}

\end{table}

In absolute terms, the number of such companies was estimated at 49,404 (39,994, 61,864), which is~a~significant part of the Polish economy. Given that this is~the first study in Poland attempting to estimate the prevalence of informal employment in Poland based on company level administrative data, this figure cannot be compared with past results. To put our findings in the context of what is~currently known about the phenomenon, they can be compared with existing direct estimates of informal activities in Poland, which indicate that 7.4\% of economically active people reported working without a~formal agreement \citep{Balance2017}, while according to survey results produced by Statistics Poland, this level is~lower: only 4.5\% of the total number of workers are employed  without a~written contract \citep{CSO2015}. The  Eurobarometer survey (2013) indicates that 3\% of employees carried out  undeclared paid activities, in addition to regular employment. \citep{EuropeanCommission2014}. In contrast, the newest indirect estimates suggest that the shadow economy accounts for 11.4\% of Poland's GDP \citep{Dybka2019}. Given that direct measures tend to be underestimated  and taking into account numerous biases related to indirect estimates, as well as discrepancies between these methods, further methodological development is~probably required. Because our method is~able to overcome many of the biases related to survey studies, it can provide a~new reliable measurement of the shadow economy in Poland, and could be applied to other economies.

We also checked the stability of our results and the selection of instrumental variables by applying the following sensitivity analysis. For the model selected based on information criteria we calculated 8 new models with different sets of variables for the selection and outcome model. Table \ref{tab-sens-analysis} presents results of the robustness checks. 

\begin{table}[ht!]
\centering
\small
\caption{Sensitivity analysis of estimates of the target parameter based on different combinations of variables in the selection and outcome equation for the generalized Heckman's sample selection (probit-logit)  model with the Gumbel copula}
\label{tab-sens-analysis}
\begin{tabular}{llrr}
  \hline
 Selection & Outcome & $\widehat{\overline{Y}}_{\text{prev}}^{\text{GH}}$ [in \%] & AIC\\
  \hline
  ind, size, complaints, re(distr) & ind, size, s(unemp) & 5.67 & 249,473 \\
  \hline
  ind, size, complaints, re(distr) & ind, size, s(unemp),  re(distr) & 33.97  & 248,208\\
  ind, size, complaints & ind, size, s(unemp),  re(distr) & 33.98  & 256,921\\
  ind, size,  re(distr) & ind, size & 4.00  & 249,562\\
  ind, complaints, re(distr) & ind, s(unemp) & 5.80  & 251,383\\
  ind, complaints, re(distr) & ind, size, s(unemp) & 5.71  & 251,273\\
  size, complaints, re(distr) & ind, size, s(unemp) & 5.38  & 254,812\\
  ind, size, complaints, re(distr) & size, s(unemp) & 12.46  & 249,606\\
  ind, size, complaints, re(distr) & ind, s(unemp) & 18.78 & 249,697\\
   \hline 
\end{tabular}
\begin{flushleft}
\textit{Note}: \texttt{ind} -- industry (NACE rev2), \texttt{size} -- company size, \texttt{re(distr)} -- random effect for district, \texttt{complaints} -- number of complaints (at district level), \texttt{s(unemp)} -- spline registered unemployment (at district level).
\end{flushleft}

\end{table}

Whether the district random effect is~included in both selection and outcome models or only in the outcome model has no effect on the estimated prevalence of informal work. This suggests that we correctly specified \texttt{district} as an instrumental variable for the selection mechanism. Removing another instrumental variable, i.e. \texttt{complaints}, does not change the results significantly. On the other hand, the exclusion of two variables, i.e. NACE section and company size, from the outcome model leads to an increase in the prevalence of informal employment to over 12\% and 18\%.

\section{Conclusion}\label{summary} 

The problem of informal employment is~gaining more and more attention from researchers and policy makers. However, because of the clandestine and heterogeneous nature of this phenomenon, methodological approaches used to measure the prevalence of informal activities are still in need of further development. Existing direct methods based on survey data, which tend to measure the lower bound of informal employment, are very rare owing to high implementation costs. Moreover, given the sensitive nature of the subject, survey responses may be dishonest and give biased results.

Against this background, in this study, we aimed to estimate the prevalence of informal employment in Poland using administrative data. To do this, we employed company level data from inspections conducted by the National Labour Inspectorate (PIP) in 2016. We used data about 25,901 Polish enterprises that were active mainly in the wholesale and retail trade, repair of motor vehicles and motorcycles and construction. The majority of audited entities are micro and small companies, employing up to 49 workers. To ensure the representatives of our results we combined the data from labour inspections with administrative records about active payers of social security contributions  provided by the Polish Social Insurance Institution (ZUS). In this way we obtained a~large, reliable data source of informal employment activities in Polish registered enterprises.

In order to estimate the level of informal activities in Poland we adopted the definition used by PIP, where informal employment is~defined as employment without a~formal contract or where an employee is~not registered for mandatory social security. Our methodology involved the use of a~bivariate probit-logit sample selection model assuming a~non-Gaussian (copula) correlation between errors from the selection and outcome model associated with the purposive sample.

We estimated that the problem of informal employment occurred in almost 6\% of Polish registered enterprises. This figure cannot be compared with previous results because of the lack of comparable company level studies. However, reviewing past evidence on informal employment in Poland, it can me assumed that our estimates are reliable. Moreover, as far as we know, ours is~the first study of informal employment to make use of administrative data, which makes our estimates unique and is~an incentive to continue research in this field.


\bibliographystyle{chicago}

\bibliography{bibl}

\clearpage

\appendix

\begin{center}
{\large\bf APPENDIX}
\end{center}

\section{Details about the National Labour Inspectorate and the The Polish Social Insurance Institution}\label{source-desc}

\subsection{Structure of the National Labour Inspectorate}

\begin{figure}[ht!]
    \centering
    \includegraphics[width=0.8\textwidth]{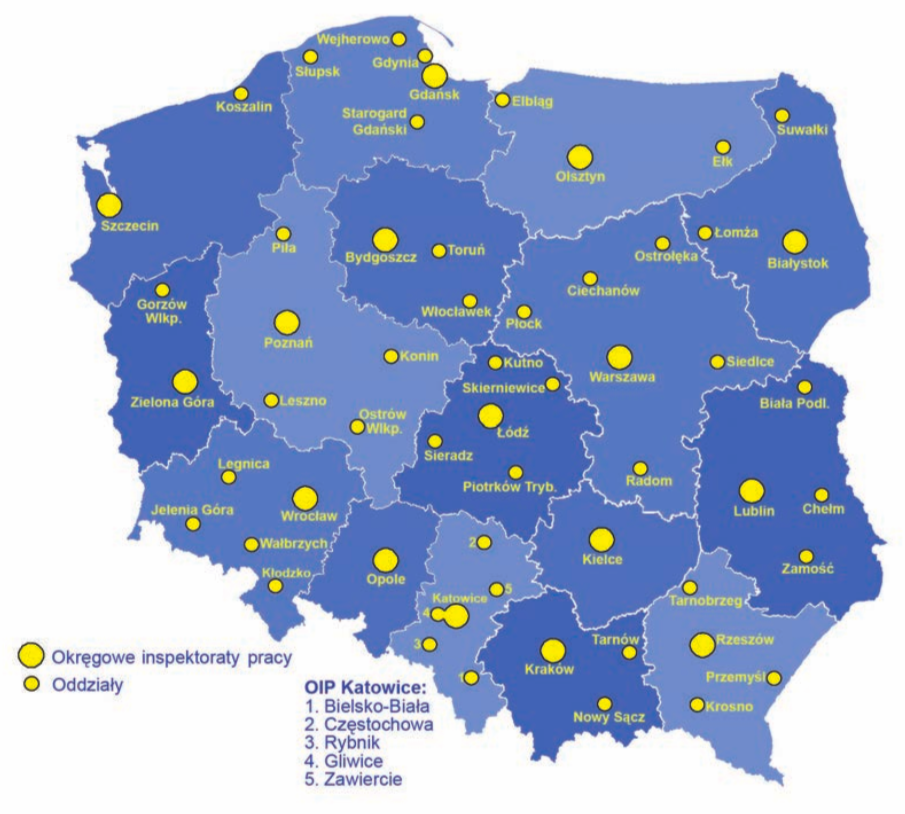}
    \caption{Territorial structure of the National Labor Inspectorate. Big circles denote District Labor Inspectorates (16) and small circles branches (42)}
    \label{nli-structure}
\end{figure}

PIP is~accountable to the Parliament and operates by virtue of the Act of 13 April 2007 on the National Labour Inspectorate. PIP consists of the Chief Labour Inspectorate and 16 District Labour Inspectorates. Inspections are administered by the Chief Labour Inspector with the assistance of deputies.

PIP verifies compliance with the labour code and inspects all employers and entrepreneurs who are not employers but who make use of work performed by natural persons for their benefit, irrespective of the legal basis on which such work is~performed. PIP is~an agency established to enforce compliance with labor law, in particular regulations and rules regarding health and safety at work, as well as the legality of employment. 

Labour inspectors have the right to conduct, without prior notice, on-site inspections to check compliance with the labour law in all economic entities where work is~executed by natural persons. In cases of violations of employees' rights, labour inspectors have the power to impose a~fine or to lodge a~complaint with a~labour court in order for  the offender to be sanctioned. Therefore, data collected by PIP is~expected to be of high quality since it is~based on the labour law and involves cases of true non-compliance with the labour regulations.

Employees audited by PIP are linked to the main head office of inspected companies, not to their the place of residence. Therefore, estimates based on these data cannot be directly compared to existing studies of the informal sector in Poland e.g. sample surveys. 

\subsection{Data preparation}\label{data-prep}

The prevalence of informal work was estimated as follows. First, we processed the raw data to correct erroneous records (i.e. companies where the number of informal workers was higher than the number of audited employees) and missing data concerning location (4 records), NACE codes (5 records) and the number of employees (8 records). Second, we disregarded sections B, D and O because of the low number of companies. Third, we adjusted the number of employees to meet the definition used in the ZUS data. The PIP data contained information about the number of employees based on the REGON register, which was outdated. Taking into account the number of audited employees we updated the actual number of employees. As a result of data processing the  initial number of companies was reduced from 26,261 to 25,901.

\begin{table}[ht!]
\centering
\caption{First six rows and selected columns from the data set used for the modelling procedure}
\label{tab-sample-data}
\begin{tabular}{lllllrr}
  \hline
District & Industry & Size & Selected & Informal & N & $\ldots$ \\ 
  \hline
1 & A & to 9 & No & No &  20  & $\ldots$\\ 
  1 & A & to 9 & Yes & No &   5  & $\ldots$\\ 
  1 & A & 10-49 & No & No &   5  & $\ldots$\\ 
  1 & F & to 9 & No & No & 188  & $\ldots$\\ 
  1 & F & to 9 & Yes & No & 1  & $\ldots$\\ 
  1 & F & to 9 & Yes & Yes &   3  & $\ldots$\\ 
  $\ldots$ & $\ldots$ & $\ldots$ & $\ldots$ & $\ldots$ &   $\ldots$ \\ 
   \hline
\end{tabular}
\end{table}

Finally, as we did not have access to unit-level data with identifiers, we decided to concatenate data from PIP and ZUS. The output file with selected records and columns containing data used for the estimation is~presented in Table \ref{tab-sample-data}. Each row represents companies with a certain combination of characteristics (\texttt{variables}): location (\texttt{District}), NACE section (\texttt{Industry}; 16 levels), number of employees (\texttt{Size}; 4 codes: to 9, 10-49, 50-249, over 250), whether a company with a certain combination of variables was included in the PIP sample (\texttt{Selected}) and whether it employed at least one person illegally (\texttt{Informal}). Column $N$ indicates the number of companies with a given combination of characteristics; all rows sum up to the size of the population. For instance, the last three records show that within the group of companies located in district 1, conducting activity classified into section F and employing up to 9~employees, 188~were not selected for inspection, 4~were selected and 3~of them employed at least one person informally. 

\subsection{Definition of the contribution payer}\label{payer-def}

A payer of social insurance and/or health insurance contributions is~a~legal person, an organizational unit without legal personality or a~natural person who is~obliged to pay contributions for insured persons, and in the case of individuals, also for themselves. Thus, contribution payers include entities paying remuneration for work or services provided on the basis of membership in agricultural production co-operatives, or under contracts of mandate, contracts for specific tasks, agency agreements, out-of-service contracts and other types of contracts involving payment of remuneration. The category of contribution payers also includes social welfare centers - with respect to persons receiving social assistance benefits, district labor offices - in relation to the unemployed, the Social Insurance Institution - in relation to persons receiving maternity benefits or benefits in the amount of a maternity benefit (definition according to the The Polish Social Insurance Institution).

\clearpage

\section{Details about the National Labour Inspectorate}

\subsection{Descriptive statistics}\label{app-nli-det}

\begin{table}[ht!]
\centering
\small
\caption{Distribution of the number of employees in the population (ZUS), selected (audited) companies and companies that employed informally}
\begin{tabular}{lrrr}
  \hline
Number of employees & Population & Selected & Informal \\ 
  \hline
 to 9 & 80.45 & 70.52 & 66.28 \\ 
  10-49 & 15.56 & 24.78 & 28.96 \\ 
  50-249 & 3.40 & 3.59 & 3.35 \\ 
  250+ & 0.59 & 1.11 & 1.42 \\ 
   \hline
  Number obs & 871 327 & 25 901 & 8 546\\
\end{tabular}
\end{table}

\begin{table}[ht!]
\centering
\small
\caption{Distribution of companies by section in the population (ZUS), selected (audited) companies and companies that employed informally}
\begin{tabular}{lrrr}
  \hline
Industry (NACE section) & Population & Selected & Informal \\ 
  \hline
  A -- Agriculture, Forestry and Fishing  & 1.31 & 1.56 & 1.29 \\ 
  C -- Manufacturing & 11.76 & 17.56 & 16.60 \\ 
  E -- Water supply & 0.63 & 0.83 & 0.74 \\ 
  F -- Construction & 11.74 & 14.42 & 13.91 \\ 
  G -- Wholesale and retail trade & 27.88 & 34.03 & 32.76 \\ 
  H -- Transportation and storage & 6.78 & 4.40 & 4.86 \\ 
  I -- Accommodation and food service activities & 3.72 & 7.33 & 8.68 \\ 
  J -- Information and communication & 2.24 & 1.26 & 1.44 \\ 
  K -- Financial and insurance activities & 1.98 & 0.76 & 0.81 \\ 
  L -- Real estate activities & 3.33 & 1.28 & 1.33 \\ 
  M -- Professional, scientific and technical activities & 8.88 & 3.68 & 3.70 \\ 
  N -- Administrative and support service activities & 3.00 & 3.62 & 4.79 \\ 
  P -- Education & 4.53 & 1.85 & 1.79 \\ 
  Q -- Human health and social work activities  & 4.62 & 2.11 & 2.48 \\ 
  R -- Arts, Entertainment and recreation  & 1.72 & 0.97 & 0.87 \\ 
  S -- Other service activities & 5.88 & 4.35 & 3.96 \\ 
   \hline
\end{tabular}
\end{table}

\clearpage

\section{Detailed description of the model}\label{app-model-desc}

\subsection{Selection of the link function}

\begin{table}[ht]
\centering
\scriptsize
\caption{Comparison of link functions for the selection and outcome variable}
\label{tab-link-funs}
\begin{tabular}{llrrr}
  \hline
    Selection \& Outcome model & Copulae & BIC & AIC & $\widehat{\overline{Y}}_{\text{prev}}^{\text{GH}}$ [in \%] \\ 
  \hline
 probit \& logit & G0 & 253,035 & 249,473 & 5.67 \\ 
  probit \& logit & AMH & 253,039 & 249,475 & 4.74 \\ 
  probit \& probit & G0 & 253,039 & 249,477 & 5.75 \\ 
  probit \& probit & AMH & 253,043 & 249,480 & 4.83 \\ 
  probit \& logit & N & 253,042 & 249,481 & 7.36 \\ 
  probit \& logit & J0 & 253,046 & 249,483 & 3.10 \\ 
  probit \& logit & C0 & 253,044 & 249,483 & 11.55 \\ 
  probit \& probit & N & 253,046 & 249,485 & 7.51 \\ 
  probit \& probit & J0 & 253,048 & 249,485 & 3.14 \\ 
  probit \& probit & C0 & 253,047 & 249,487 & 11.79 \\ 
  logit \& logit & G0 & 253,077 & 249,500 & 5.87 \\ 
  logit \& probit & G0 & 253,081 & 249,504 & 5.99 \\ 
  logit \& logit & AMH & 253,084 & 249,506 & 5.02 \\ 
  logit \& logit & N & 253,083 & 249,507 & 7.56 \\ 
  logit \& logit & C0 & 253,083 & 249,507 & 11.46 \\ 
  logit \& probit & AMH & 253,088 & 249,510 & 5.13 \\ 
  logit \& probit & N & 253,087 & 249,511 & 7.73 \\ 
  logit \& probit & C0 & 253,087 & 249,511 & 11.74 \\ 
  logit \& logit & J0 & 253,094 & 249,516 & 3.37 \\ 
  logit \& probit & J0 & 253,096 & 249,518 & 3.41 \\ 
   \hline
\end{tabular}
\end{table}

\clearpage
\subsection{Description of the final model}\label{model-details}

\begin{table}
\scriptsize
\caption{Parameter estimates for the selection and outcome equation in the Generalized Heckman's sample selection model}
\label{tab-models}
\begin{center}
\begin{tabular}{ll c c }
\hline
&& Selection & Outcome \\
\hline
\multicolumn{4}{c}{\textbf{Fixed Effects}} \\
\hline
\hline
 (Intercept) & & $-2.08 \; (0.05)^{***}$    & $-1.01 \; (0.11)^{***}$ \\
Industry
& Manufacturing             & $0.11 \; (0.02)^{***}$     & $0.06 \; (0.12)$        \\
& Water supply; sewerage, waste management and remediation activities             & $-0.00 \; (0.04)$          & $0.01 \; (0.19)$        \\
& Construction            & $0.07 \; (0.02)^{**}$      & $0.14 \; (0.12)$        \\
& Wholesale and retail trade; repair of motor vehicled and motorcycles            & $0.07 \; (0.02)^{**}$      & $0.15 \; (0.12)$        \\
& Transportation and storage             & $-0.22 \; (0.03)^{***}$    & $0.33 \; (0.13)^{**}$   \\
& Accomodation and food service activities            & $0.29 \; (0.03)^{***}$     & $0.44 \; (0.12)^{***}$  \\
& Information and communication             & $-0.23 \; (0.03)^{***}$    & $0.28 \; (0.16)$        \\
& Financial and insurance activities             & $-0.43 \; (0.04)^{***}$    & $0.26 \; (0.19)$        \\
& Real estate activities             & $-0.41 \; (0.03)^{***}$    & $0.21 \; (0.16)$        \\
& Professional, scientific and technical activities             & $-0.36 \; (0.03)^{***}$    & $0.15 \; (0.13)$        \\
& Administrative and support service activities             & $0.06 \; (0.03)^{*}$       & $0.53 \; (0.13)^{***}$  \\
& Education             & $-0.55 \; (0.03)^{***}$    & $0.02 \; (0.15)$        \\
& Human health and social work activities             & $-0.44 \; (0.03)^{***}$    & $0.39 \; (0.14)^{**}$   \\
& Arts, Entertainment and recreation             & $-0.33 \; (0.03)^{***}$    & $-0.00 \; (0.18)$       \\
& Other service activities             & $-0.15 \; (0.03)^{***}$    & $0.07 \; (0.13)$        \\
Size
& 10-49           & $0.30 \; (0.01)^{***}$     & $0.33 \; (0.03)^{***}$  \\
& 50-249           & $0.16 \; (0.02)^{***}$     & $-0.04 \; (0.07)$       \\
& over 250           & $0.39 \; (0.03)^{***}$     & $0.26 \; (0.12)^{*}$    \\
Complaints  &   ratio  & $5.54 \; (1.56)^{***}$     &                         \\
\hline
\multicolumn{4}{c}{\textbf{Random Effects}} \\
\hline
& s(district)         & $354.67 \; (371.00)^{***}$ &                         \\
& s(unemployment) &                            & $8.87 \; (8.99)^{***}$  \\
\hline
\multicolumn{4}{c}{\textbf{Summary statistics}} \\
\hline
& AIC                 & 217183.96                  & 32416.77                \\
& BIC                 & 220229.69                  & 32612.17                \\
& Log Likelihood      & -108217.31                 & -16180.52               \\
& Deviance            & 216434.62                  & 32361.03                \\
& GCV score           & 3.26                       & 2.96                    \\
& Num. obs.           & 25063                      & 8192                    \\
\hline
\multicolumn{3}{l}{\scriptsize{$^{***}p<0.001$, $^{**}p<0.01$, $^*p<0.05$}}
\end{tabular}
\end{center}
\end{table}

\end{document}